# Stabilization of antiferromagnetism in CeFe$_2$ alloys: Effects of chemical and hydrostatic pressure

Arabinda Haldar[1], K. G. Suresh[1,*], A. K. Nigam[2], A. A. Coelho[3] and S. Gama[3]
[1]Department of Physics, Indian Institute of Technology Bombay, Mumbai – 400076, India
[2]Tata Institute of Fundamental Research, Homi Bhabha Road, Mumbai 400005, India
[3]Instituto de Fisica 'Gleb Wataghin,' Universidade Estadual de Campinas-Unicamp, CP 6165, Campinas 13 083 970, SP, Brazil

E-mail: suresh@phy.iitb.ac.in

**Abstract.** Effects of Al, Mn and Sb dopings in CeFe$_2$ and effect of applied pressure have been investigated. Al doping gives rise to the FM-AFM transition and a reduction in the magnetic moment and T$_C$ values, clearly indicating the growth of the AFM component. Mn and Sb dopings only cause a reduction in T$_C$ value. It is found that in general external pressure enhances the antiferromagnetism in both the pure and the doped alloys. Enhancement of the Ce *4f*- Fe *3d* hybridization as a result of dopings and with the external pressure may be the reason for the stabilization of antiferromagnetism in these alloys.

[*]Author to whom any correspondence should be addressed.

## 1. Introduction

Magnetism of pure and doped $CeFe_2$ has attracted considerable interest from the researchers for many decades [1]-[3]. The strong hybridization between Ce 4*f* and Fe 3*d* electronic states makes $CeFe_2$ very special among the series of $RFe_2$(R= Rare earth) compounds. Although Ce is a light rare earth, 4*f* electrons hybridize antiferromagnetically with the 3*d* electrons. This is due to the quenching of orbital 4*f* moment by band formation [1]. Lattice parameter, Curie temperature and the magnetic moment show a strong deviation from the normal trend observed in the $RFe_2$ series [1, 4]. As shown previously, the ground state of $CeFe_2$ at ambient pressure can be viewed as a canted ferromagnet (FM), i.e., a small antiferromagnetic (AFM) component (<0.1$\mu_B$) superimposed on a dominant ferromagnetic component [5].

It was predicted that with the application of pressure, the magnetic moments of Fe and Ce would decrease rapidly. Theoretical calculations have shown that with a compression of the unit cell by about 6-7%, the compound loses most of its magnetic property, thereby becoming paramagnetic [6]. It was experimentally found that at 220 kbar and at low temperatures $CeFe_2$ loses its magnetism [7]. Nevertheless, it is clear that this critical pressure is considerably lower than that estimated for $YFe_2$ (1050 kbar [8]). Further comparison with $YFe_2$ shows that, at low pressures, the Curie temperature increases with pressure for $YFe_2$ [9], whereas it decreases in $CeFe_2$ despite a continuous decrease of the iron moment in both the compounds. This again indicates that $CeFe_2$ occupies a special position amongst the $RFe_2$ compounds [7].

Enhancement of antiferromagnetic spin correlations has been shown in $CeFe_2$ single crystals at pressures up to 8 kbar [10]. A recent pressure study on $CeFe_2$ single crystals shows that the transition to the AFM state at low temperatures occurs at pressures less than 20 kbar [11]. They have also found that the temperature at which it undergoes the antiferromagnetic transition (at the Neel temperature, $T_N$) increases monotonically with pressure. Neutron diffraction studies indicate that the static Fe magnetic moment is suppressed to be only half of that under P=1 bar at 5K and that the antiferromagnetic spin fluctuation with a propagation vector q=$(1/2,1/2,1/2)$ is enhanced by applying hydrostatic pressure of 15 kbar [12].

On the other hand, it is also well known that, at ambient pressure too, the stabilization of the antiferromagnetic component in $CeFe_2$ can be achieved with the help of substitution of small amounts (3-6%) of selected elements such as Co, Al, Ga, Ru, Ir, Os and Re [2, 3, 13]. The first order nature of the FM-AFM transition (on cooling) gives rise to distinct features of supercooling/superheating and kinetic arrest behavior [14]. It is worth noting that, among these substitutions, while Al causes an increase in the lattice parameter, Co results in lattice contraction. However, surprisingly in both these cases, the AFM spin correlations get enhanced, giving rise to the stable low temperature AFM state. Therefore, it is clear that the chemical pressure (positive or negative) alone is not responsible for the stabilization of the AFM state. Effect of external pressure has been reported on a few substituted $CeFe_2$ compounds as well [15, 16].

The moments on the trivalent Ce ion and Fe in the $ReFe_2$ series are $2.54\,\mu_B$ and $1.77\,\mu_B$, respectively [17]. From the polarized neutron study of $CeFe_2$, Ce and Fe moments are found to be $-0.14\,\mu_B/atom$ and $1.17\,\mu_B/atom$, respectively [18]. Due to hybridization of Ce 4$f$ and Fe 3$d$ electronic states, Ce remains in mixed valance state (+3.29) rather than its normal trivalent state in $CeFe_2$ and this must be the reason for the anomalous magnetic moments of Ce as well as Fe in $CeFe_2$. Therefore, it is of interest to investigate the effect of pressure created by chemical substitution (chemical pressure) and by external means (hydrostatic pressure) on the magnetic properties. With the view of comparing the pressure effects in different doped $CeFe_2$ compounds which undergo lattice expansion and contraction as a result of substitutions, we report the pressure studies on Al, Mn and Sb doped $CeFe_2$. The compounds studied in this work are $CeFe_2$, $Ce(Fe_{1-x}Al_x)_2$ [x = 0.01, 0.025 & 0.05], $Ce(Fe_{0.95}Mn_{0.05})_2$, $Ce(Fe_{0.95}Sb_{0.05})_2$. While Al and Mn doping are known to cause lattice expansion, Sb causes lattice contraction.

**2. Experimental Details**

Polycrystalline $CeFe_2$, $Ce(Fe_{0.95}Mn_{0.05})_2$, $Ce(Fe_{0.95}Sb_{0.05})_2$ and $Ce(Fe_{1-x}Al_x)_2$ [x = 0.01, 0.025 and 0.05] compounds have been prepared by arc melting the stoichiometric proportion of the constituent elements of at least 99.9% purity in water cooled copper hearth in argon atmosphere. The resulting ingots were turned upside down and remelted several times to ensure homogeneity. The as-cast samples were annealed for 10 days in the sequence: 600 ºC for 2 days, 700 ºC for 5 days, 800 ºC for 2 days and 850 ºC for a day [2]. The structural characterization of these annealed compounds was done by room temperature powder x-ray diffraction (XRD) using Cu-K$_\alpha$ radiation. XRD data was refined by Rietveld refinement method using *Fullprof* suite program. Structural parameters have been calculated from this refinement. Magnetization measurements have been performed in a Vibrating Sample Magnetometer attached to a Physical Property

Measurement System (PPMS, Quantum Design Model 6500) and/or a SQUID magnetometer. The magnetization measurements under applied pressures have been carried out using a Cu-Be clamp type cell attached to the SQUID magnetometer. The maximum pressure that can be obtained in the cell was 10kbar. Magnetization has been measured in zero field cooled (ZFC), field cooled cooling (FCC) and field cooled warming (FCW) modes. The sample was in the form of a small piece weighing a few tens of milligrams and of irregular shape.

**3. Results and Discussions**

**$Ce(Fe_{1-x}Al_x)_2$ [x = 0, 0.01, 0.025 & 0.05]**

Room temperature x-ray diffraction patterns along with the Rietveld refinement plots confirm the phase purity of all the compounds. All of them possess the cubic structure with space group: $Fd\bar{3}m$. The lattice parameters are found to increase as Al concentration increases from 0 to 0.05.

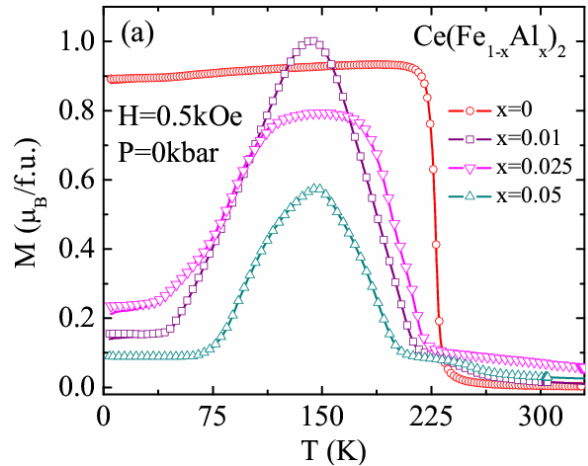

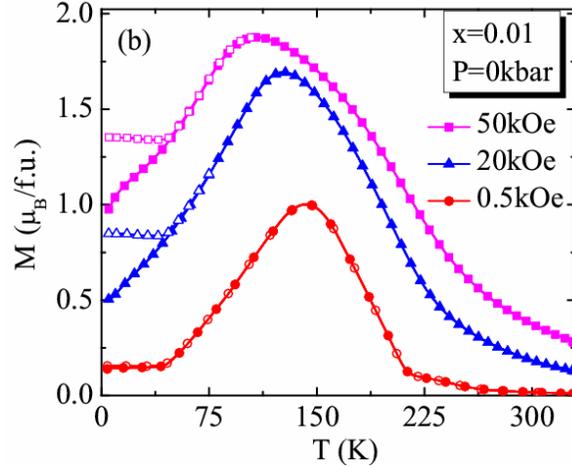

**Figure 1**. (a) Temperature dependence of the magnetization of $Ce(Fe_{1-x}Al_x)_2$ compounds for x= 0, 0.01, 0.025, 0.05. (b) Temperature dependence of ZFC and FCW magnetization of $Ce(Fe_{0.99}Al_{0.01})_2$ in different applied fields. Open and closed symbols refer to FCW and ZFC modes of data respectively.

Figure 1(a) shows the temperature variation of magnetization data at a low field of H=500 Oe for pure and Al doped samples at ambient pressure. While parent compound, $CeFe_2$ shows a normal ferromagnetic behavior with a Curie temperature ($T_C$) of 230K [19], Al doped compounds show the stabilized AFM state at low temperatures, in agreement with earlier reports [2, 20]. There is a significant reduction in $T_C$ with substitution of Al along with an increase in the FM-AFM transition temperature, $T_N$. Both these observations point towards the fact that the AFM strength increases with Al. In figure 1(b), magnetization data is shown for x=0.01 in different fields. The effect of external field is to reduce the AFM coupling and favor a FM behavior in the material. With increase in field, the $T_C$ values increase and the $T_N$ values decrease. The difference between ZFC and FCW data increases with increase in field, which is consistent with our recent results on Ga substituted $CeFe_2$ [19]. It is clear from figure 1(b) that the AFM phase is very strong at 0.5kOe, but the strength decreases as the field is increased to higher values. This gives rise to the bifurcation between the ZFC and the FCW data at 20 kOe and 50 kOe.

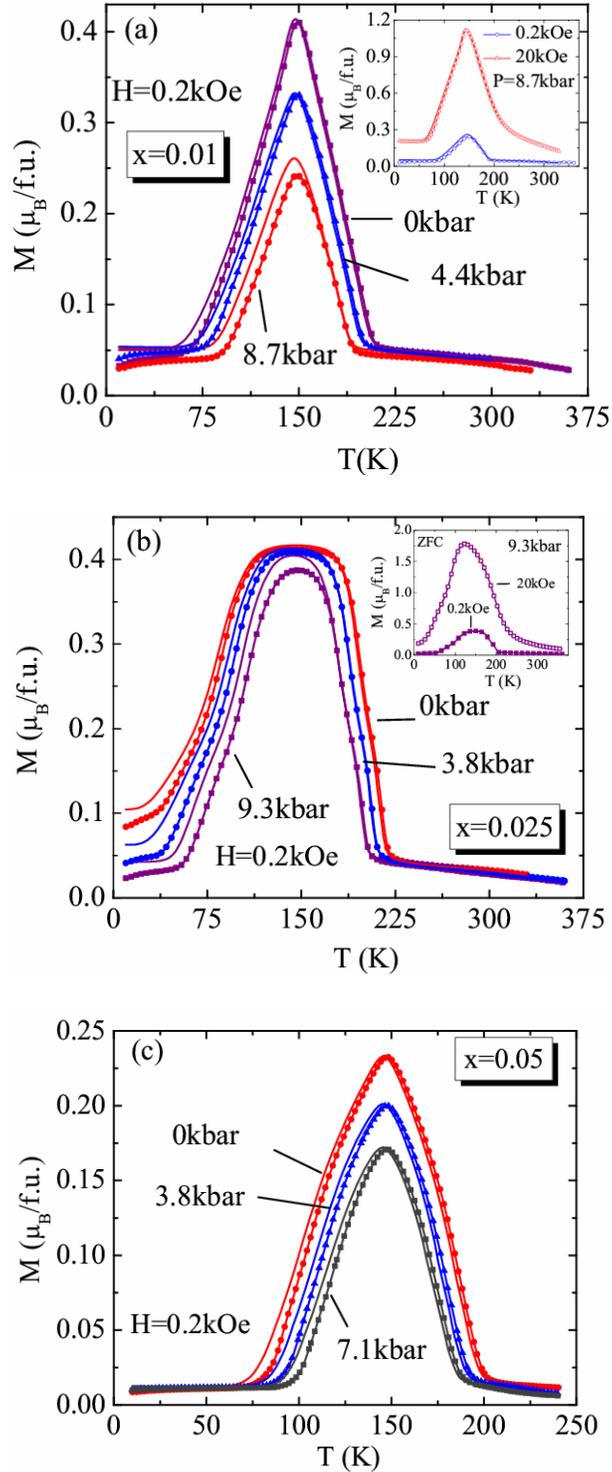

**Figure 2**. Temperature variation of magnetization in ZFC and FCC modes under various applied pressures for the compounds (a) x=0.01, (b) x=0.025, and (c) x=0.05. In each plot filled symbols refer to ZFC and solid lines refer to FCC data. The insets in (a) and (b) are the M-T curves at two different fields at the highest pressure.

Figure 2(a)-(c) shows the temperature dependence of magnetization at different pressures for Ce(Fe$_{1-x}$Al$_x$)$_2$ compounds with x=0.01, 0.025 and 0.05. In all the three compounds, with increase in pressure the T$_C$ decreases and the T$_N$ increases, which suggests that the effect of external pressure is to stabilize the AFM phase. The magnetization values also decrease with pressure. These observations are due to the pressure induced enhancement of antiferromagnetic correlations in the system. With pressure the widening of Ce $f$ level occurs which in turn enhances hybridization between Ce 4$f$ and Fe 3$d$ electronic states. As a result, the Fe–Fe exchange interaction responsible for the Curie temperature, gets weaker and the outcome is a decrease of T$_C$ and an increase of T$_N$. Another important observation from figure 2 is that the difference between ZFC and FCC data increases with the increase in pressure. This can be understood in the following way. The FM-AFM transition in these compounds is known to be accompanied by a structural transition (at T$_N$) from cubic to rhombohedral structure [2]. Because of this magnetostructural coupling, increase of pressure prefers the rhombohedral phase. Therefore, the magnetization variation during cooling (FCC) and heating (ZFC) is expected to be different. This is because the energy barrier that the system encounters during heating and cooling will be different. This explains the considerable ZFC-FCC difference at higher pressures, compared to that at the ambient pressure.

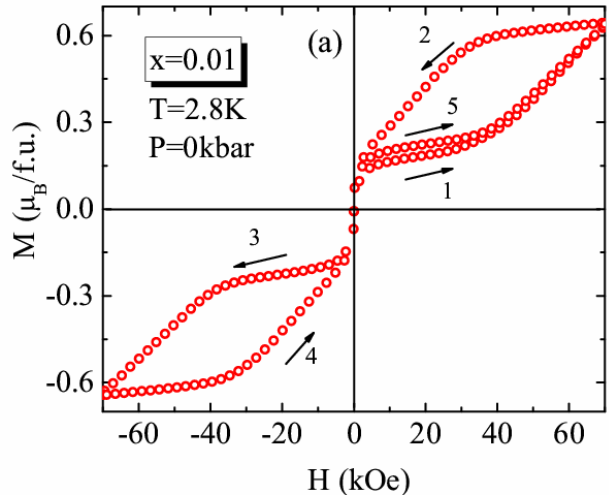

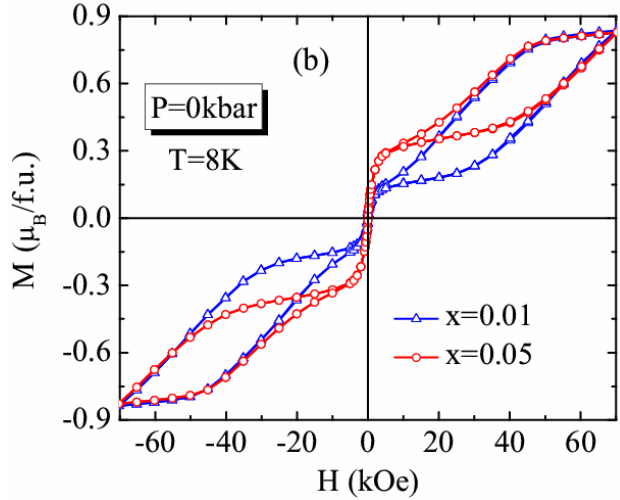

**Figure 3**. (a) M-H isotherms at T=2.8 K at ambient pressure for $Ce(Fe_{0.99}Al_{0.01})_2$. Arrows indicate the direction of field in which data have been taken. (b) *M-H* loops at T=8 K for x=0.01 and 0.05

Figure 3(a) shows the five loop *M(H)* isotherm at T=2.8 K under ambient pressure (P=0 kbar) for x=0.01 compound. This measurement has been made with a field sweep rate of 100 Oe per second. The sample was zero-field cooled from room temperature (paramagnetic region) to the measurement temperature. With increasing field, the moment shows a rapid initial increase, then becomes flat and shows a smooth jump at a critical field $H_c \sim 28kOe$. The moment does not get saturated even in a field of 70kOe. Between increasing and decreasing cycles, a large hysteresis is observed, but with zero coercivity. This observation confirms the first order nature of the magnetic field induced AFM to FM transition. Another observation to be noticed here is that the virgin curve remains outside of the envelope curve which was also observed earlier for x=0.04 [20]. Super cooling and kinetic arrest associated with the first order transition are responsible for this behavior [20]. With increase in the concentration of Al, the antiferromagnetic correlation increases, as reflected by the increase in the ciritical field values shown in figure 3(b). Although Al substitution causes increase in the lattice parameter, it shows the effect similar to that with applied hydrostatic pressure. It may be due to modification in density of sates (DOS) around Fermi level with Al substitution. Another point to be noted in is that at 8K the $5^{th}$ loop follows the same path as the $1^{st}$ loop, which is not the case at T=2.8K (see figure 3(a)). This indicates that thermal fluctuations reduce the effects associated with the kinetic arrest of the FM phase.

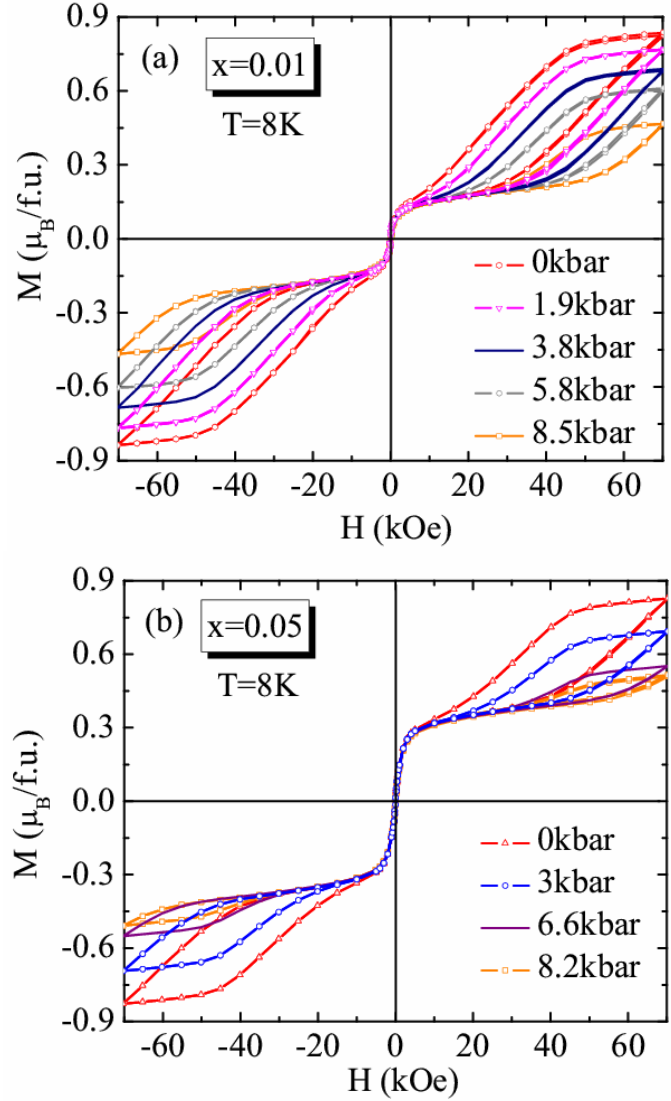

**Figure 4**. M *vs*. H loop at T=8 K under various pressures for (a) x=0.01 and (b) x=0.05 compounds.

In figure 4(a)-(b), the nine-loop magnetization isotherms are shown for x=0.01 and 0.05 at 8 K and at various pressures. For this measurement also, the sample was zero field cooled from the paramagnetic region to 8 K and then the data was taken in increasing and decreasing modes for both positive and negative fields. For a given pressure, there is no difference between the data in the first and second cycles of magnetic field variation. The M-H behavior is almost identical to that seen in figure 3(b). Here also, the moment does not get saturated at a field of 70kOe. With pressure the critical field ($H_c$) increases and the moment reduces considerably. Effect of pressure seems to reinforce the 3*d*-4*f* hybridization present in Al doped $CeFe_2$ compounds which in turn reinforces the AFM correlations in the material. This gives rise to an increase in $H_C$ as shown in
9

figure 5. Comparison of 8 K and 50 K data in figure 5 implies that thermal fluctuation converts the system towards more parallel alignment of moments, reducing the $H_C$ values. The decrease in moment seen in figure 4 can also be attributed to the enhanced hybridization with pressure.

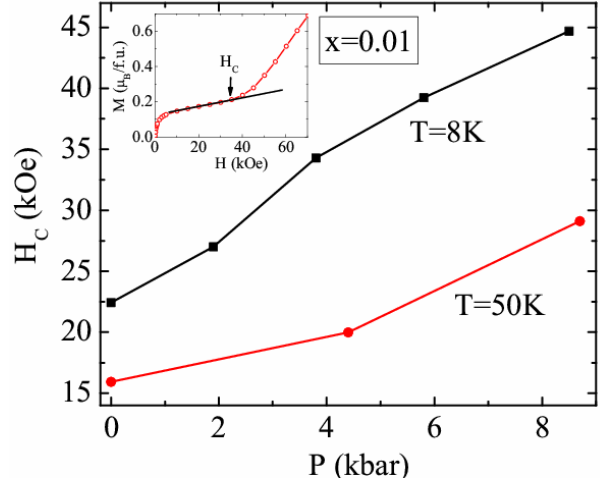

**Figure 5**. Pressure dependence of critical field in $Ce(Fe_{0.99}Al_{0.01})_2$ at different temperatures.

It can be noticed that the AFM-FM transition in Al doped compounds is rather smooth, unlike the case in Ru, Re and Ga doped compounds [19, 21]. Furthermore, the region of existence of ferromagnetism is very narrow (see figures 1 & 2) in the Al-doped case as compared to other substitutions. Similarly, the M-H isotherms at temperatures below $T_N$ in the case of Ru, Re and Ga doped compounds are generally accompanied by multiple sharp jumps. Such jumps are absent in the case of Al doped compounds. Kennedy et al. have shown that there is a broad overlap between two magnetic phases (AFM and FM) in the case of Al doping compared to the narrow overlap seen with other dopants [3], which may be the reason for the smooth transition in the former. The application of magnetic field at low temperatures favors the FM phase, and as the field is increased, the volume ratio of the FM phase to the AFM phase evolves gradually in this case, due to a broad overlap region between these phases. So instead of getting sharp jumps in the magnetization isotherms, a smooth behavior is found across the field induced AFM-FM transition.

**Mn and Sb doped CeFe$_2$ compounds**

At the ambient pressure, Mn doped CeFe$_2$ compounds are known to be ferromagnets [22, 23]. Likewise, we find that Sb substitution also does not stabilize the AFM fluctuations. Even a pressure of about 9 kbar is not able to stabilize the AFM phase at low temperatures in both these compounds. However, there is a monotonic reduction in the ferro-para transition temperature, ($T_C$) in both the cases, as shown in figure 6, which also shows the variation in pure CeFe$_2$. As observed in the earlier section, pressure has a considerable effect on the hybridization between the Ce 4$f$ and Fe 3$d$ electronic states. It is clear from figure 6 that the pressure dependence of $T_C$ is maximum in CeFe$_2$, whereas it is lower in both Mn and Sb doped compounds. In fact, the pressure dependence seen in these two cases is found to be nearly equal to that of Al doped compounds. Therefore, it is clear that the effect of applied pressure is maximum in CeFe$_2$, as there is no chemical pressure in the undoped case. Like in Al doped compounds, the pressure seems to enhance the 3$d$-4$f$ hybridization in Mn and Sb substituted compounds to such an extent that there is a reduction in $T_C$. However, this hybridization is not enough to result in a strong AFM state as seen in the Al doping.

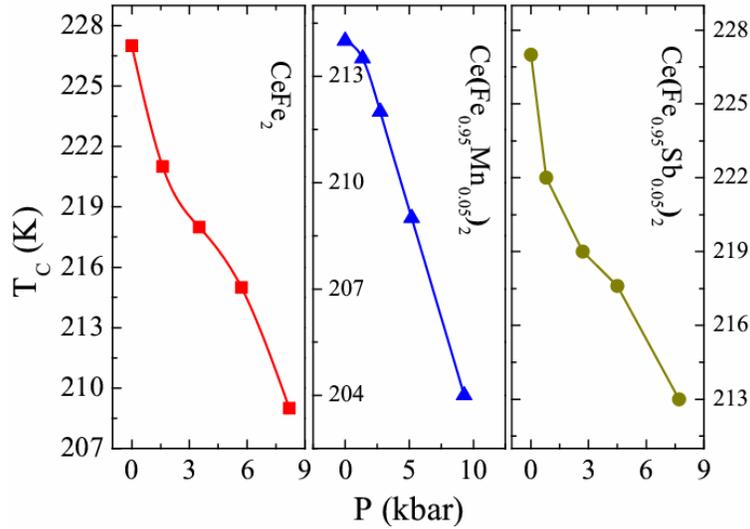

**Figure 6**. Variation of Cuire temperature ($T_C$) with external pressures for undoped, Mn and Sb doped CeFe$_2$ compounds.

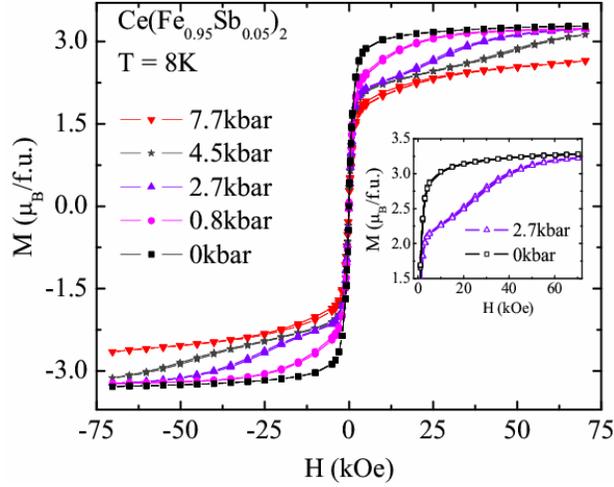

**Figure 7**. Isothermal magnetization loops at T=8K for Ce(Fe$_{0.05}$Sb$_{0.05}$)$_2$. The inset shows the first quadrant of the M-H loop at the ambient pressure and at 2.7 kbar.

At ambient pressure, the saturation moments are 2.6 and 2.7 $\mu_B/f.u.$ for CeFe$_2$ and Ce(Fe$_{0.95}$Mn$_{0.05}$)$_2$ respectively at 8 K. With the application of pressure no significant change is observed in these values. On the other hand, at ambient pressure, in Ce(Fe$_{0.95}$Sb$_{0.05}$)$_2$, the moment is 3.3 $\mu_B/f.u.$ which reduces to 2.6 $\mu_B/f.u.$ at the highest pressure. Another interesting observation in the case of Ce(Fe$_{0.95}$Sb$_{0.05}$)$_2$ is the metamagnetic transition found at P= 2.7 kbar (see inset of figure 7). At higher pressures, a similar transition seems to appear at fields above 75 kOe. The presence of metamagnetic transition indicates that the system possesses a AFM state, at least in very small fraction, at low temperatures, though the M *vs*. T curve does not show any clear evidence of this fact. However, this tendency is in agreement with the reduction in T$_C$ seen in figure 6.

Therefore, it is clear that there is an enhancement in the AFM component in the case of Mn and Sb dopings in CeFe$_2$, though a transition to the AFM state is not observed. This may be due to the fact that the 4*f*-3*d* overlap increases only nominally in the case of Mn and Sb dopings or these substitutions may be unable to change the DOS near Fermi level in such a way to stabilize AFM state in these materials.

## 4. Conclusions

We have shown the pressure effect on the magnetization of the undoped and doped $CeFe_2$ compounds. It is found that in general external hydrostatic pressure enhances the AFM component in doped $CeFe_2$. The increase in the $T_N$ values and the decrease in the magnetic moment and $T_C$ values clearly show the growth of the AFM component. The results have been discussed on the basis of the mixed valent behavior of Ce ion as well as the change in the hybridization between the Ce 4$f$ and Fe 3$d$ electronic states. External pressure seems to enhance the hybridization present in these materials. Results obtained on different dopings indicate that it is not the ionic radii, but the extent of hybridization and modification in the mixed valency that are responsible for the stabilization of the antiferromagnetic phase. Detailed band structure calculations on all these dopings and experimental data using techniques like photoelectron spectroscopy are essential to unravel the underlying physics of the anomalies observed in different dopings. The effect of external pressure observed here opens up the opportunity to theoretically model these materials.

*Acknowledgement*

KGS and AKN thank BRNS for the financial support for carrying out this work.